\documentclass[manuscript]{emulateapj}
\usepackage{graphics}
\usepackage{color}
\begin{document}

\newcommand{\kms}{\mbox{km~s$^{-1}$}}
\newcommand{\s}{\mbox{$''$}}
\newcommand{\mloss}{\mbox{$\dot{M}$}}
\newcommand{\my}{\mbox{$M_{\odot}$~yr$^{-1}$}}
\newcommand{\ls}{\mbox{$L_{\odot}$}}
\newcommand{\ms}{\mbox{$M_{\odot}$}}
\newcommand{\rsun}{\mbox{$R_{\odot}$}}
\newcommand\mdot{$\dot{M}$}
\def\arcdeg{\hbox{$^\circ$}}
\def\farcs{\hbox{$.\!\!^{\prime\prime}$}}
\def\gtrsim{\mathrel{\hbox{\rlap{\hbox{\lower4pt\hbox{$\sim$}}}\hbox{$>$}}}}
\def\lesssim{\mathrel{\hbox{\rlap{\hbox{\lower4pt\hbox{$\sim$}}}\hbox{$<$}}}}

\title{An Extreme High-Velocity Bipolar Outflow in the Pre-Planetary Nebula IRAS\,08005-2356}
\author{R. Sahai\altaffilmark{1} \& N.A. Patel\altaffilmark{2}
}

\altaffiltext{1}{Jet Propulsion Laboratory, MS 183-900, California
Institute of Technology, Pasadena, CA 91109}

\altaffiltext{2}{Harvard-Smithsonian Center for Astrophysics, Cambridge}

\email{raghvendra.sahai@jpl.nasa.gov}

\begin{abstract} We report interferometric mapping of 
the bipolar pre-planetary nebula IRAS\,08005-2356 (I\,08005) with an angular-resolution of $\sim$1{''}--5{''}, 
using the Submillimeter Array (SMA), in the $^{12}$CO J=2--1, 3--2, 
$^{13}$CO J=2--1 and 
SiO J=5--4 (v=0) lines. Single-dish observations, using the SMT 10-m, were made in these lines as well 
as in the CO J=4--3 and SiO J-6--5 (v=0) lines. The lines profiles are very broad, showing 
the presence of a massive ($>0.1$\,\ms), extreme high-velocity outflow ($V\sim200$\,\kms)
directed along the nebular symmetry axis derived from the HST imaging of this object. 
The outflow's scalar momentum far exceeds that available from radiation pressure of the central post-AGB star, 
and it may be launched from an accretion disk 
around a main-sequence companion. We provide indirect evidence for such a disk from its  
previously published, broad H$\alpha$ emission profile, which we propose results from   
Ly$\beta$ emission generated in the disk followed by Raman-scattering in the innermost 
regions of a fast, neutral wind. 
\end{abstract}
\section{INTRODUCTION}
Following the ejection of half or more of their mass via isotropic, slowy expanding winds, AGB stars evolve into 
planetary nebulae (PNe), which, surprisingly show a diverse range of aspherical (e.g., bipolar and multipolar) morphologies 
(e.g., Sahai \& Trauger 1998, Sahai, Morris \& Villar 2011a). Studies of pre-Planetary nebulae (PPNs), objects in transition between the AGB and
planetary nebula (PN) evolutionary phases, are critical for characterising the physical processes 
responsible for this dramatic transformation. Sahai \& Trauger (1998) proposed that 
fast collimated outflows or jets, operating during the PPN and/or very late AGB phase, are the primary agents for the 
dramatic change in the mass-loss geometry and dynamics during the AGB-to-PN evolutionary phase. However, the physical mechanism for producing these fast
outflows remains a mystery. High-angular-resolution interferometric (sub)millimeter-wave 
observations are the best way to quantitatively probe the fast outflow's dynamics and energetics -- 
crucial information for theoretical models (e.g., Akashi \& Soker 2013), and detailed numerical hydrodynamical simulations  
(e.g., Lee \& Sahai 2003, Balick et al. 2013), for PN shaping.

Such observations have resulted in the discovery of a handful of 
``extreme-outflow" PPNs -- objects in which the molecular outflows reach speeds in excess of 
$\sim$100\,\kms, e.g., Boomerang Nebula (Sahai et al. 2013), IRAS\,22036+5306 (Sahai et al. 2006), 
IRAS\,19374+2359 (S{\' a}nchez Contreras et al. 2013), IRAS\,16342-3814 (Imai et al. 2012) and HD\,101584 (Olofsson et al. 2015). 
Detailed studies of 
such extreme objects are likely to provide the best motivation for, and most stringent tests of, 
theoretical models to explain their origin (e.g., Blackman \& Lucchini 2014).

In this Letter, we report (sub)millimeter-line observations of the PPN IRAS\,08005-2356 (I\,08005), 
which clearly reveal it to be an extreme-outflow PPN. Early CO J=2--1 observations by Hu et al. (1994) resulted in 
the marginal detection of a weak, broad line. 
I\,08005's F5\,Ie central star V510\,Pup (Slijkhuis
et al. 1991) may have made a 
recent transition from ejecting oxygen-rich material to carbon-rich material
(Bakker et al. 1997). Its morphological classification is Bo*(0.55) (Sahai et al. 2007),  
i.e., it has a bipolar morphology (resolved via HST imaging, Ueta et al. 2000) with lobes 
open at their ends and a central star seen at 0.55\micron. Its optical spectrum reveals the presence of 
a prominent H$\alpha$ emission line with very broad wings (FWZI$\sim 2400$\,\kms) and a P-Cygni type blue-shifted 
absorption feature (S{\' a}nchez Contreras et al. 2008: SCetal08; also Slijkhuis et al. 1991, Klochkova \& Chentsov 2004). 
Its estimated distance ranges between 2.85\,kpc (Oppenheimer et al. 2005: OBS05) and 3--4\,kpc (Klochkova \& Chentsov 2004); we adopt 
a value of 3\,kpc.

\section{OBSERVATIONS}
The 1.3\,mm and 0.87\,mm interferometric observations were obtained with the Submillimeter Array (SMA\footnote{The Submillimeter Array is a joint project
between the Smithsonian Astrophysical Observatory and the Academia Sinica
Institute of Astronomy and Astrophysics and is funded by the Smithsonian
Institution and the Academia Sinica}) at 
Mauna Kea, Hawaii. Bandpass calibration was performed using observations
of 3C279. At 1.3 (0.87)\,mm, complex gain calibration was obtained from observations
of the quasars 0750+125 and 0730-116 (0747-331 and 0826-225), and flux calibration was obtained from observations
of Callisto (Europa). Additional 
observing parameters are listed in Table\,\ref{obs-parm}. 

SMA data were calibrated using the MIR-IDL package\footnote{http://www.cfa.harvard.edu/$\sim$cqi/mircook.html}, and images were made using the  
Miriad software. Data cubes were obtained with a velocity
resolution smoothed to 10\,\kms~per channel (to increase the S/N in each
channel). Natural weighting was used to produce all images.

Single-dish observations of the $^{12}$CO J=2--1, 3--2 and 4--3, $^{13}$CO J=2--1, and SiO (v=0) J=5--4 and 6--5 line emission
were obtained at SMT during 2014, Nov/Dec and 2015 Jan. Telescope pointing was frequently checked on VY\,CMa, 
and is estimated to be better than a small fraction of the beam. The weather was generally good, 
with system temparatures in the range $T_{sys}\sim200-250$\,K at 1.3\,mm and $T_{sys}\sim 1300$\,K at 0.7 and 0.8\,mm. 
Linear baselines were subtracted from the spectra shown. We have assumed  
main-beam efficiency factors of 0.76, 0.66, and 0.62 for the 230.6, 345.8, and 461.0\,GHz observations (Edwards et al. 2014).

The optical images were taken with the NASA/ESA Hubble Space Telescope in GO programs ID 6364 (PI: M. Bobrowsky) 
and 6366 (PI: S. Trammell), using the PC on WFPC2, and 
extracted from the Hubble Legacy Archive\footnote{A collaboration 
between the Space Telescope Science Institute (STScI/NASA), the Space 
Telescope European Coordinating Facility (ST-ECF/ESA) and the         
Canadian Astronomy Data Centre (CADC/NRC/CSA)}.

\section{RESULTS}
\subsection{Optical Imaging}
\begin{figure}[htbp]
\vskip -2in
\includegraphics[width=20cm]{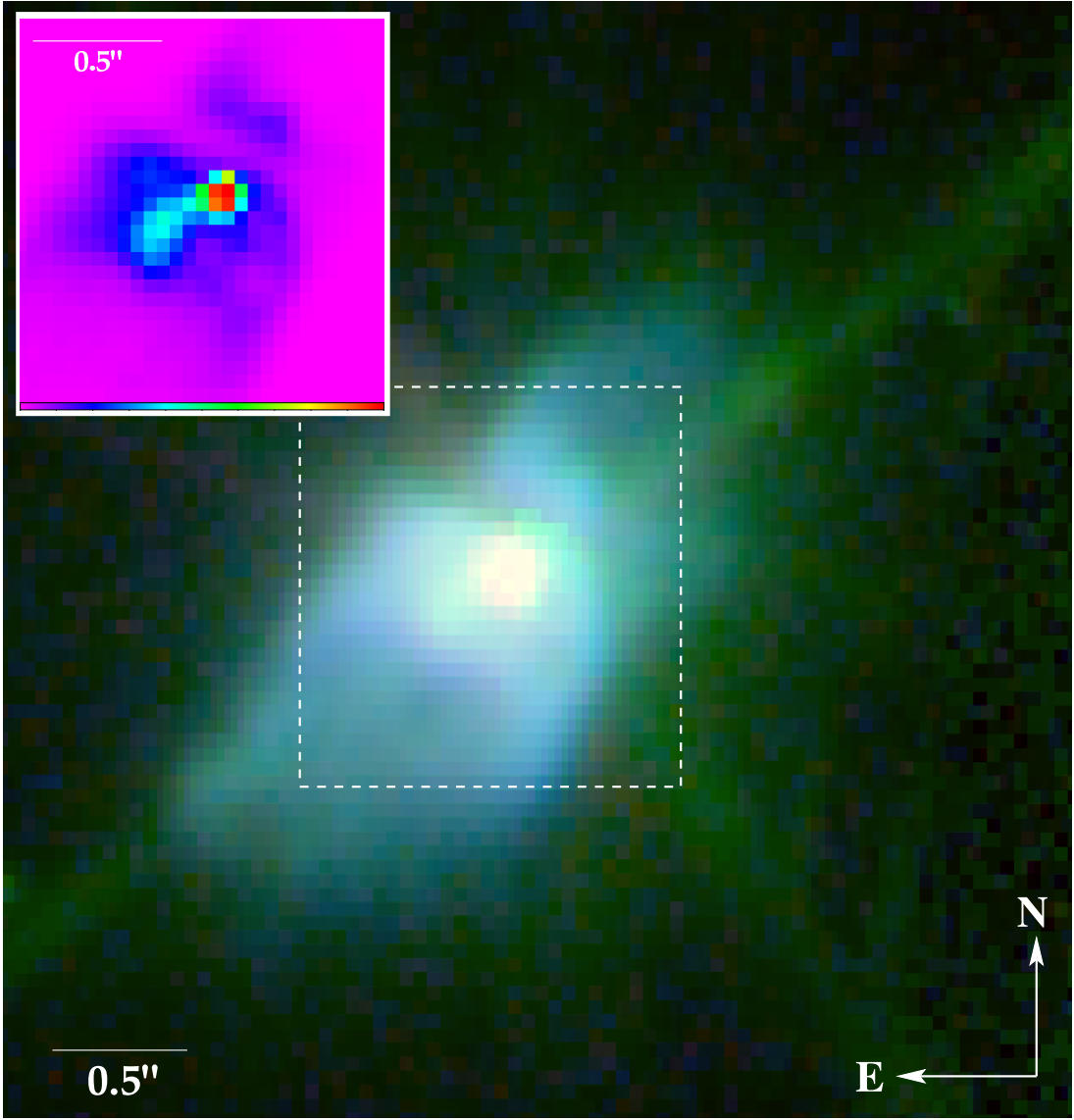}
\caption{HST color-composite (red: F675W, green: F555W, blue: F439W) image of I\,08005 (log stretch). Panel size 
is $3\farcs94\times3\farcs94$. The diagonal linear features are diffraction spikes due to the bright central source in the relatively long-exposure 
F555W image. Inset is a false-color view (linear stretch: colorbar shows counts/s from 0 to 210) of the central $1\farcs4\times1\farcs4$ region (dashed box) in F439W, highlighting the curved jet-like feature emanating from the central source.
}
\label{i08005hst}
\end{figure}

I\,08005 shows an hourglass morphology in the HST F439W, F555W, and F675W images (Fig.\,\ref{i08005hst}) -- 
the lobes, separated by a narrow dark lane, flare out from the central star's location, and attain a cylindrical shape. The SE lobe is significantly brighter than the NW one. 
A narrow, slightly-curved feature of linear extent 0\farcs38, is seen extending from the central star in the SE lobe (see inset). The feature is oriented at a $PA=120\arcdeg$, 
slightly different from the PA of the nebula symmetry axis, $143\arcdeg$. The total linear extent of the nebula along its long 
axis, as seen in the F555W image, is $\sim2\farcs6$ ($1.2\times10^{17}$\,cm).
\subsection{Millimeter-Wave Observations}
\begin{figure}[htbp]
\begin{center}
\vskip -3in
\includegraphics[width=18cm]{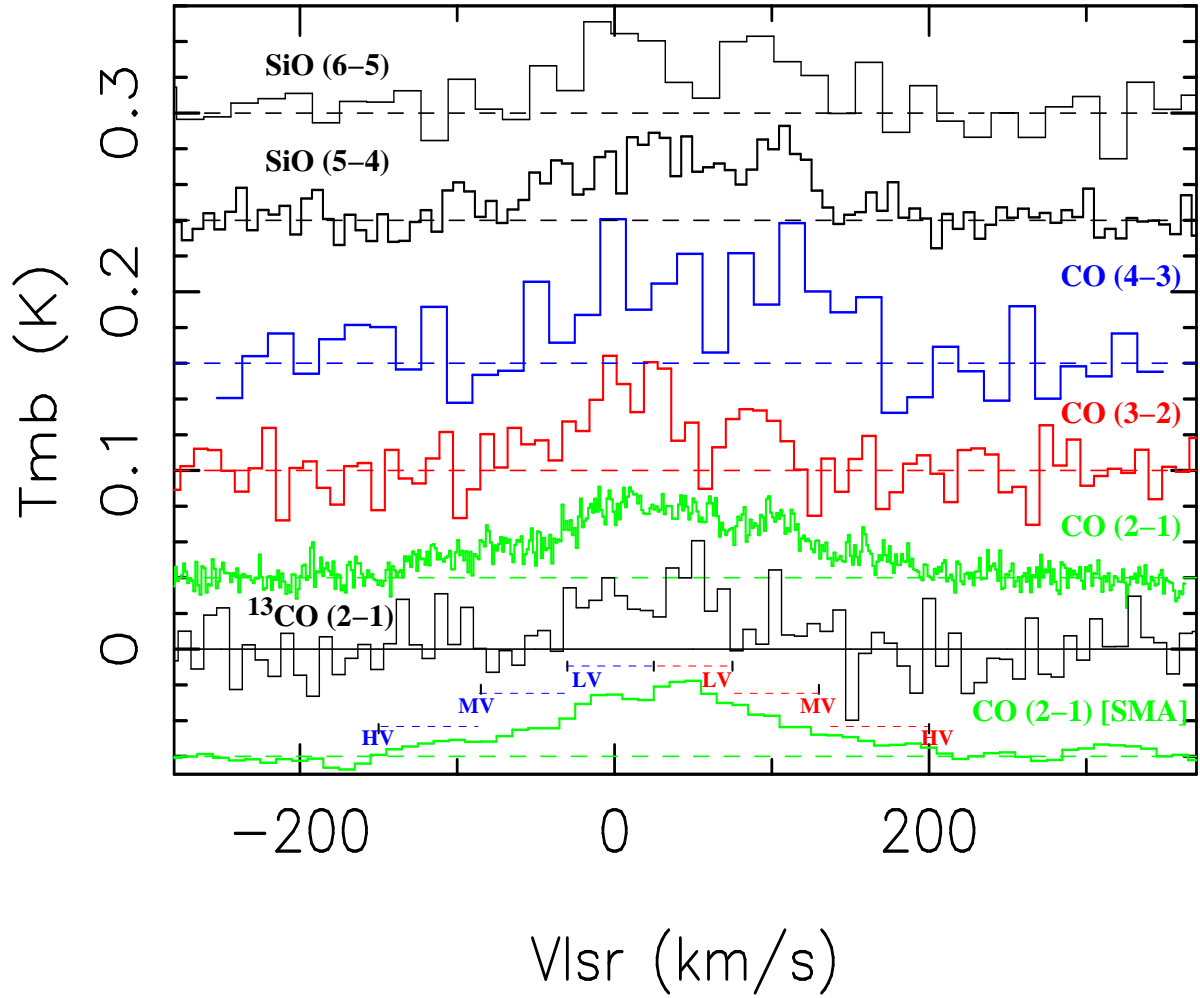}
\end{center}
\caption{Molecular-line emission from I\,08005. The SMT spectra (shown above $T_{mb}=0$ axis) have been shifted vertically for clarity, 
after scaling as follows: SiO J=6--5 and J=5--4\,($\times4$), $^{12}$CO J=4--3 and 2--1\,($\times2$), \& 
$^{13}$CO J=2-1\,($\times10$). The velocity ranges of the HV, MV, \& LV components are marked in the SMA panel. The SMA $^{12}$CO J=2--1 spectrum, shifted below $T_{mb}=0$, shows the 
total flux divided by the SMT conversion factor of 44.5 Jy/K, then scaled $\times2$, for comparison with the 
SMT spectrum. The apparent difference in the SMT and SMA CO J=2-1 line-profiles at 
$V_{lsr}\sim100$\,\kms~is probably due to non-linear baseline structure that is not removed by the linear baseline subtraction.
}
\label{cosmt}
\end{figure}

We detected the CO J=2--1, 3--2 and SiO J=5--4 lines with the SMA. The CO J=3--2 and SiO J=5--4 have significantly lower signal-to-noise, 
and the $^{13}$CO J=2--1 is tentatively detected. 
The (spatially) integrated SMA CO J=2--1 line
profile (using a $15{''}$ diameter circular aperture) shows a very broad profile 
(Fig.\,\ref{cosmt}) covering a total velocity extent of about 350\,\kms, i.e., from $-150\,\lesssim V_{lsr}\,(\kms)\,\lesssim 200$. 
The SMT CO J=2--1 line profile (Fig.\,\ref{cosmt}) is similar to the SMA one (a narrow 
emission feature at $V_{lsr}=36$\,\kms due to the presence of an unrelated line-of-sight interstellar cloud, has been removed from the 
profile by interpolation). A Gaussian fit to the profile gives a central velocity $V_{lsr}\sim25$\,\kms, and a FWHM of 196\,\kms.

\begin{figure}[htbp]
\begin{center}
\vskip -0.7in
\includegraphics[width=18cm]{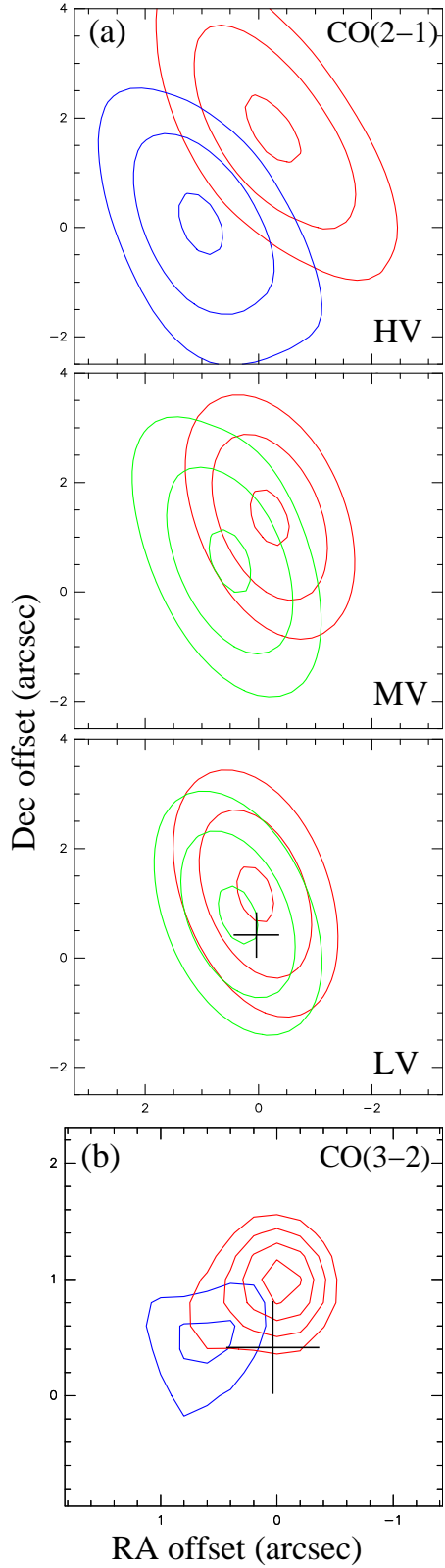}
\end{center}
\caption{SMA CO emission integrated over different velocity ($V_{lsr}$) ranges (a) J=2--1, ({\it top}) blue: $-150,-85$, red: $130,200$; 
({\it middle}) blue: $-85,-30$, red: $70,130$; ({\it bottom}) blue: $-30,25$, red: $25,75$. (b) J=3--2, blue: $-90,20$, red: $30,130$. Contour levels for CO 
J=2--1 (3--2) are: minimum=50 (46)\%, step=10 (15)\% of the peak intensities.
The central source in the HST image (marked with a cross), is offset (0\farcs041,0\farcs4) relative to the phase-center (at 0,0 with J2000 coordinates RA=08:02:40.7, Dec=-24:04:43.0), and consistent with the bipolar-outflow center, within the 
combined 1-\,$\sigma$ uncertainty, $\sim$($\pm$0\farcs4,$\pm$0\farcs4), of the absolute HST and SMT astrometry.}
\label{blured}
\end{figure}

We have divided the velocity range spanned by the CO J=2--1 line into 
three intervals on each side of the central velocity -- the extreme high-velocity or EV component (blue: $-150<V_{lsr}\,(\kms)<-85$, 
red: $130<V_{lsr}\,(\kms)<200$), a medium-velocity or MV component (blue: $-85<V_{lsr}\,(\kms)<-30$, red: $70<V_{lsr}\,(\kms)<130$), and 
a low-velocity LV component (blue: $-30<V_{lsr}\,(\kms)<25$, red: $25<V_{lsr}\,(\kms)<75$). 
The SMA CO J=2--1 emission in the blue and red-shifted parts of the EV, MV, and LV components show separations of 
about $2\farcs1$, $1\farcs1$, $0\farcs5$, respectively, along the nebular symmetry axis (Fig.\,\ref{blured}a).

These results imply the presence of a fast, bipolar outflow with a (roughly) linear velocity-gradient, 
directed along the nebular symmetry axis, and a linear extent of $\sim2\farcs1$. Since AGB mass-loss is typically spherical, with outflow velocities 
of 10--15\,\kms, we conclude that the outflow 
observed in CO J=2--1 emission is a fast, collimated post-AGB outflow, typical of the PPN evolutionary phase. We made similar plots 
from the SMA data cubes for CO J=3--2 (Fig.\,\ref{blured}b) and SiO J=5--4 (not shown), 
dividing the emission velocity-range into two halves only (blue: $-90<V_{lsr}\,(\kms)<20$, 
red: $30<V_{lsr}\,(\kms)<130$) due to the lower S/N in these 
lines, and found that these show a separation of $0\farcs85$ roughly along the 
nebular axis. This separation is equal to that derived from our CO J=2--1 data for the same red and blue outflow-velocity ranges, 
implying that the SiO outflow's linear extent is likely comparable to that of CO.

We also detected the CO J=4--3, J=3--2 and $^{13}$CO J=2--1 as well as the  
SiO (v=0) J=5--4 and 6--5 lines with the SMT, but with lower S/N (Fig.\,\ref{cosmt}). 
We have compared the total CO J=2--1 line fluxes from our SMA and SMT data, as these have the highest 
S/N, and find that these are consistent within 15\%, i.e., within 
typical calibration uncertainties) -- we conclude that the SMA data do not suffer 
from any significant flux losses.

Continuum images were obtained after
removing spectral regions containing line emission, using the 
Miriad task uvlin. 
No continuum emission was detected from I\,08005. Using line-free channels, we 
find a $1\sigma$ noise of 1.05 (1.2) mJy/beam in the USB (LSB) at 1.3\,mm and 11 (18) mJy/beam in the USB (LSB) at 0.87\,mm.

\section{The Post-AGB Bipolar Outflow}

\subsection{Outflow Properties}
We now determine the physical properties of the high-velocity bipolar outflow (e.g., scalar momentum, mass, and age) from an analysis 
of the molecular-line data.
The observed ratios of the CO J=2--1, 3--2 and J=4--3 fluxes (in K\,\kms) are $1:1.27:1.2$, 
which imply corresponding non-beam diluted flux ratios of $1:\rm R_{32/21}:\rm R_{43/21}=1:0.6:0.3$, 
since the source is unresolved by the SMT beams of $32{''}$, $22{''}$, and $16{''}$ in these lines.

We first make a simple emission model 
assuming a spherical outflow at a uniform excitation temperature equal to the kinetic temperature, $\rm T_{kin}$, and carry 
out a least-squares fit to the line fluxes\footnote{derived using the formulation in Olofsson et al.\,(1990)}, varying $\rm T_{kin}$, the outer radius ($\rm R_{out}$), and mass of the emitting region ($\rm M_g$). 
We find $\rm R_{out}=1.4{''}$ and $\rm T_{kin}=13.6$\,K. 
However, the derived $\rm T_{kin}$ is rather low compared to energy of 
the CO J=4 level (55.4\,K), and modeling with the non-LTE RADEX code  (van der Taak et al. 2007) shows that the 
J=4--3 excitation temperature, $\rm T_{ex(43)}$, is lower than $\rm T_{kin}$ by about (20--25)\% 
due to the relatively low CO line optical depths in our model ($<0.01$) and
the average density, $\rm n_{av}\sim3.5\times10^4$\,cm$^{-3}$, implied by the emitting region's mass and volume. 
Thus, the resulting, corrected 
model $R_{43/21}\sim0.11$, too low to fit the data.

Using RADEX, we find that the average density required to bring $\rm T_{ex(43)}$ closer to $\rm T_{kin}$ in order to fit 
the observed source brightness temperature ratios, is $\rm n_{av}>7.5\times10^4$\,cm$^{-3}$; the required 
$\rm T_{kin}=15.5$\,K, giving a 
source size of $2\farcs4$, comparable to the CO source-size estimated 
from the SMA data. The model source size is similar to the size derived from the HST image, implying that 
the CO emission likely comes from the dense walls of the bipolar lobes seen in scattered light. 
The CO column density is, 
$N_{CO}>5\times10^{17}$\,cm$^{-2}$ (and the J=2--1, 3--2, and 4--3 lines have 
optical depths of 0.48, 0.44, 0.20). Assuming that the average emitting column 
is equal to the source radius, we find a CO-to-H$_2$ abundance ratio of, $f_{CO}=1.3\times10^{-4}$, in 
reasonable agreement with the value typically assumed for PPNs, $2\times10^{-4}$ (e.g., Bujarrabal et al. 2001). Assuming a spherical emitting volume, the mass is, 
$\rm M_g=0.076$\,\ms. The $^{12}$CO/$^{13}$CO abundance ratio is, f($^{12}$C/$^{13}$C)=9.6, from fitting the  
$^{12}$CO/$^{13}$CO J=2--1 line flux ratio corrected for the different beam-dilutions (8.1). 

The above values of $\rm M_g$ and f($^{12}$C/$^{13}$C) 
are lower limits since this is our ``minimum-mass" model --  
models with higher values of $\rm n_{av}$ are allowed. However, the total mass increases more slowly than the average density, since 
in models with higher values of $\rm n_{av}$, the emitting region's size is smaller. 
For example, if we use the dust mass, $\rm M_d$=0.0019\,\ls\,$(D/2.85\,\rm kpc)^2$ derived 
by OBS05 from a detailed 2D radiative transfer model of I\,08005's SED, scale it to D=3\,kpc, and 
adopt a typical gas-to-dust ratio for oxygen-rich 
AGB stars, $\rm M_g/M_d=200$, we get $\rm M_g=0.42$\,\ms. For this value of $\rm M_g$, we need a 
CO model with $\rm n_{av}\sim8\times10^5$\,cm$^{-3}$, a factor $\sim10$ higher than in 
our minimum-mass model; $\rm T_{kin}=11.5$\,K, and $N_{CO}=1.6\times10^{18}$\,cm$^{-2}$.  
Since the CO J=2--1 optical depth is higher ($\tau _{21}=1.95$), the abundance ratio is higher,  f($^{12}$C/$^{13}$C)=17.

We calculate the scalar momentum using the formulation 
described in Bujarrabal et al. (2001). Using our minimum-mass model, we find $P_{sc}\sim2.8\times10^{39}$\,g\,cm\,s$^{-1}$ for an
inclination angle of the nebular axis to the sky-plane, $i=30\arcdeg$ (OBS05). 
The kinetic energy in the outflow is $E_{kin}\sim2.6\times10^{45}$\,erg. 
These values of $P_{sc}$ and $E_{kin}$
lie near the upper end of 
the range for PPNs, $10^{37-40}$ g\,cm\,s$^{-1}$ and $10^{42-46}$\,erg (Bujarrabal et al. 2001). 
This outflow cannot be driven by radiation
pressure because the lobes' dynamical (expansion) time-scale, $\rm t_{dyn}=190$\,yr (from dividing the 
model CO shell-size by the CO J=2-1 FWHM line-width) is much smaller than that required by radiation
pressure to accelerate the observed bipolar outflow to its current speed, $t_{rad}=P_{sc}/(L/c)\sim 6.6\times10^4$ yr, 
given I\,08005's luminosity
of $6980$\,\ls~at D=3\,kpc (using the value derived by OBS05, $6300(D/2.85\,\rm kpc)^2$\,\ls.) The mass-loss rate in 
the outflow is $>5.8\times10^{-4}$\,\my.

The observed SMT SiO J=6--5 to 5--4 line flux ratio is $0.94$ (with about $\pm15$\% uncertainty), implying an 
intrinsic flux ratio (i.e., corrected for beam-dilution) of $\rm R(SiO)_{65/54}=0.66$, 
since the source is unresolved by the SMT beams of $34{''}$ and $28\farcs4$ in these 
lines. Using RADEX, we find that for $T_{kin}=15.5$\,K, the 
average $\rm H_2$ density in the SiO emission region is, $n_{av}>2\times10^5$\,cm$^{-3}$ in order 
to fit the observed value of $\rm R(SiO)_{65/54}$. The fractional SiO abundance 
is $\sim4\times10^{-5}$, close to the maximum value 
assuming cosmic abundances ($6\times10^{-5}$), and comparable to 
the observed maximum circumstellar SiO abundance in oxygen-rich AGB stars ($few\times10^{-5}$: Sahai \& Bieging 1993). 
If the SiO emission comes from shocked gas at a higher kinetic temperature than that for CO, the minimum density required is higher.

\subsection{A Central Accretion Disk?}
Evidence of very fast outflows in I\,08005 comes from three independent
probes: (i) CO data (presented here), (ii) 
H$\alpha$ spectrum in  SCetal08, and (iii) OH maser emission (VLA mapping) reported in 
Zijlstra et al. (2001). The OH maser features cover the velocity
range of $0<V_{lsr}\,(\kms)<100$, i.e., roughly the range covered by red half of the CO J=2--1 profile. In contrast to
OH and CO emission, the H$\alpha$ absorption probes atomic gas. Since the
H$\alpha$ absorption feature is not spatially resolved in the ground-based long-slit spectra, it is
difficult to directly establish its relationship to the fast
molecular outflow, however the close agreement between the terminal outflow velocities derived for these indicates 
that they are closely associated. 
The atomic gas may be located inside the lobes and may constitute 
unshocked material of an underlying jet, and/or the interface between
the latter and the lobe walls, as proposed for IRAS\,22036+5306 (Scetal06).
The jet-like feature seen in the HST image (Fig.\,\ref{i08005hst},\,inset) may  
represent the precessing jet's signature close to its launch-site.

The high-speed bipolar outflow in I\,08005 may be driven by an accretion disk. Bakker et al. (1997) find numerous narrow, double-peaked, chromospheric 
emission lines from neutral and singly ionized metals, and propose that these might arise in an accretion disk.
We conjecture that the broad H$\alpha$ wings seen towards this object might arise as a result of Raman-scattering of Ly$\beta$ emission 
generated by such a disk -- a process that  
produces an H$\alpha$ profile with a width that is a factor 6.4 larger than the Ly$\beta$ width and a $\lambda^{-2}$ wing
profile. The very wide ($FWZI\sim2400$\,\kms) H$\alpha$ line wings in I\,08005 show this 
shape, furthermore a weak emission feature around 6830\AA~with $FWZI\sim150$\kms, corresponding to 
Raman-scattering of the 1032\AA~component of the O\,VI doublet at $\lambda\lambda1032,1038$ is also seen (SCetal08); the 7088\AA~feature that 
corresponds to the 1038\AA~component is a factor 4 weaker, and too weak to be visible. 

Other line-broadening mechanisms include electron scattering and emission from a rotating disk. 
Arrieta \& Torres-Peimbert (2003) find that electron scattering requires extreme
densities (e.g., $n_e > 10^{12}$ cm$^{-3}$ in M\,2-9) making that an
implausible mechanism. Keplerian rotation in a disk around the central star is too low to directly account for the broad line-width in I\,08005 (and PPNe 
in general: Sahai et al. 2011b) -- 
taking the radius for the central post-AGB star of I\,08005 to be $R\sim50$\rsun~(Slijkhuis et al. 1991) and a nominal stellar mass of 1\ms, we find 
the maximum rotation speed is $V_{mx} < 56.5$\,\kms, too low for generating the extreme line-wings of the 
H$\alpha$ profile, even with the factor 6.4 increase provided by Raman-scattering. We therefore conclude that 
the accretion disk in I\,08005 must be around a much smaller star, e.g., a main-sequence companion. 

The Raman-scattering likely occurs in the innermost regions of the fast neutral wind seen via its blue-shifted absorption feature 
signature in the H$\alpha$ profile. In this case, following Sahai et al. (2011b), we 
scale from the minimum scattering column density, $N_s=10^{19-20}$ cm$^{-2}$ needed to achieve the Raman conversion efficiency necessary for 
producing the very broad observed line-widths (see Fig.\,1 of Lee \& Hyung 2000), to find that
$dM_s/dt > (0.35-3.5)\times10^{-8}$\my ($r_w/30$\,AU) (D/3\,\rm kpc) ($V_{exp}$/100\,\kms) ($N_s/10^{19}$cm$^{-2}$), where $r_w$ is the
radius of the neutral wind where the Raman-scattering occurs. $V_{exp}$ is set equal to $0.5\times FWHM$ of the broad CO J=2--1 line profile,  
and similar to the neutral outflow's average outflow-velocity derived by SCetal08. The mass-loss rate requirement is easily met in I\,08005 since 
OBS05 derive a mass-loss rate of $4.4\times10^{-6}$\my~for a collimated fast wind in the 24-780\,AU region around the central star, 

\section{Discussion}

I\,08005 is very similar to other extreme-outflow oxygen-rich PPNs, IRAS\,22036+5306 (Sahai et al. 2006) and HD\,101584 (Olofsson et al. 2015, Sivarani et al. 1999) in the properties of its collimated fast outflow (expansion-velocity, scalar-momentum, and kinetic-energy), 
and its H$\alpha$ emission profile (very broad wings, P-Cygni absorption). But in striking contrast to these objects, it 
appears to lack mm and submm continuum emission from its central region -- the ratio 
of its 1.3\,mm to 60\,micron flux is $<10^{-4}$ ($3\,\sigma$), compared to 
$0.6\times10^{-3}$ and $1.5\times10^{-3}$ for HD\,101584 and IRAS\,22036+5306, respectively. 

Both PPNs, as well as the ``disk-prominent" sub-class of young post-AGB objects (dpAGB objects: Sahai et al. 2011c), which unlike PPNs, show little or no extended nebulosity, and have central stars that are radial-velocity binaries (e.g., van Winckel et al. 2008), emit relatively strong mm/submm continuum emission from their central regions. This emission has been attributed to the presence of substantial masses of 
cool, mm-sized grains (Sahai et al. 2011c, Gielen et al. 2007, de Ruyter et al. 2006). Their origin is not understood at present but is potentially a key probe 
of important mass-ejection processes occurring during the late-AGB and post-AGB evolutionary phases, especially those that lead to the formation of 
large dusty equatorial disks or torii. For example, in HD\,101584, where a 
binary companion has been found from radial-velocity variations (Bakker et al. 1997), Olofsson et al. (2015) propose a scenarion in which 
both the formation of the central, equatorially-dense 
mass structure and the collimated outflow result from a common-envelope (CE) event -- but the latter  
does not release enough energy to drive the mass-ejection, and another mechanism augments or even dominates it. 

The apparent lack of a substantial mass of material in the equatorial waist of I\,08005 suggests that perhaps it did not undergo a CE event, 
and its collimated outflow (and collimated outflows in PPNs generally) may be launched differently, possibly from an accretion disk as we have suggested earlier.  
I\,08005 is thus a key post-AGB object for further detailed study. For example, ALMA can be used to probe its compact central region, in order 
to search for weak mm/submm continuum emission that may be present but was below our sensitivity limit, and 
for the presence of gas (and its kinematics) associated with its dusty waist. 

\acknowledgements We thank Ken Young and Ray Blundell for granting filler time on the SMA. R.S. is thankful for partial financial support
for this work from a NASA/ ADP and LTSA grant. R.S.'s contribution
to this paper was carried out at JPL, California Institute of Technology, under a contract with NASA.

\begin{table}[!t]
\caption{SMA Observational Parameters}
\label{obs-parm}
\begin{tabular}{llllllllll}
\hline
Lines          & Freq.Range & $\Delta\nu$   & Array\,(\# Ants.) & Time\tablenotemark{1}  &  Beam, PA                  & T$_{sys}$ & Epoch \\
               & [GHz]      & [kHz]         &                   & [hr]  & ${''}\times{''},\,\arcdeg$ &   [K]  & yy/mm/dd  \\
\hline
$^{12}$CO(2-1) & USB\tablenotemark{2}: 228.85--232.85 & 812.5   &   compact\,(6)     & 1  & $4.99\times2.73$,\,20.8      &  100--230   & 13/12/11  \\
continuum\tablenotemark{3}     & 228.9--229.6,\,229.9--230.2 & ...     & ...    & ... & ...   & ...  & ... \\
              &      230.8--232.9                & ...     & ...    & ... & ...   & ...  & ... \\ \\
$^{13}$CO,\,C$^{18}$O(2-1) & LSB\tablenotemark{2}: 216.85--220.85&   ...     & ...    & ... & ...   & ...  & ... \\
SiO(5-4),v=0 &  &  &  &   &  &  &  \\
continuum\tablenotemark{3}     & 216.9--217.0,\,217.3--219.4 &  ...     & ...    & ... & ...   & ...  & ...  \\
\hline
 $^{12}$CO(3-2) & LSB\tablenotemark{2}: 342.0--346.0  & 812.5   &   extended\,(7)     & 2 & $0.85\times0.56$,\,33.5      & 240--400   & 14/04/13  \\
continuum\tablenotemark{3}       & 342.0--345.4   & ...     & ...    & ... & ...   & ...  & ... \\ \\
continuum\tablenotemark{3}       & USB\tablenotemark{2}: 354.0--358.0    & ...   & ...     & ...  & ...   & ...     & ... \\
\hline
\tablenotetext{1}{On-source integration time}
\tablenotetext{2}{The frequency range for the full spectrometer passband covering line and continuum regions}
\tablenotetext{3}{The frequency ranges used for extracting the continuum}
\end{tabular}
\end{table}

\end{document}